\documentclass[a4paper]{article}
\usepackage{INTERSPEECH2022}
\usepackage{multicol,multirow}
\usepackage{url,cite}
\usepackage{hyperref}
\usepackage{xcolor}
\usepackage{tabularx,lipsum}
\hypersetup{
    colorlinks=true,
    linkcolor=purple,
    filecolor=magenta,      
    urlcolor=purple,
}
\newcommand{\newpara}[1]{\vspace{8pt}\noindent\textbf{#1}}
\newcolumntype{Y}{>{\centering\arraybackslash}X}

\title{Pushing the limits of raw waveform speaker recognition}
\name{Jee-weon Jung$^{1,*}$\thanks{$^*$Equal contribution}, You Jin Kim$^{1,*}$, Hee-Soo Heo$^1$, Bong-Jin Lee$^1$, Youngki Kwon$^1$, Joon Son Chung$^2$}
\address{
  $^1$Naver Corporation, South Korea\\
  $^2$Korea Advanced Institute of Science and Technology, South Korea
}
\email{\{jeeweon.jung, youjin.kim117\}@navercorp.com, joonson@kaist.ac.kr}
\begin{document}

\maketitle

\begin{abstract}
In recent years, speaker recognition systems based on raw waveform inputs have received increasing attention.
However, the performance of such systems are typically inferior to the state-of-the-art handcrafted feature-based counterparts, which demonstrate equal error rates under 1\% on the popular VoxCeleb1 test set. 
This paper proposes a novel speaker recognition model based on raw waveform inputs. 
The model incorporates recent advances in machine learning and speaker verification, including the Res2Net backbone module and multi-layer feature aggregation. 
Our best model achieves an equal error rate of 0.89\%, which is competitive with the state-of-the-art models based on handcrafted features, and outperforms the best model based on raw waveform inputs by a large margin.
We also explore the application of the proposed model in the context of self-supervised learning framework. 
Our self-supervised model outperforms single phase-based existing works in this line of research. 
Finally, we show that self-supervised pre-training is effective for the semi-supervised scenario where we only have a small set of labelled training data, along with a larger set of unlabelled examples.\\
\end{abstract}

\section{Introduction}
Despite the increasing popularity of end-to-end machine learning, the use of handcrafted features (such as MFCCs or mel-filterbanks) remain prevalent in many speech processing tasks~\cite{snyder2018x,desplanques2020ecapa,chung2018voxceleb2,cai2020fly,you2008svm,dehak2010front}.
However, models that directly operate upon raw waveforms are becoming more common across several speech-related fields with the recent advances in deep learning. 
The majority of the existing works in the area have used vanilla convolutional layers or parameterised filterbank layers to process raw waveform inputs~\cite{baevski2020wav2vec,sainath2015learning,ravanelli2018speaker,pariente2020filterbank}. 

The trend towards end-to-end models can also be observed in the speaker recognition literature. 
Since the first raw waveform speaker recognition model that employs a VGGNet-styled vanilla convolutional layer~\cite{jung2017d}, several studies now adopt direct modelling from raw waveforms~\cite{zhu2021vector,jung2019rawnet,li2021fdn,lin2020wav2spk,kim2021rawnext}.

However, most raw waveform speaker recognition models suffer from performance degradation compared to handcrafted feature-based counterparts.
Even the latest architecture demonstrates equal error rate (EER) of 1.29\%~\cite{kim2021rawnext}, whereas the widely adopted ECAPA-TDNN architecture and its variants have consistently reported EERs under 1\%~\cite{desplanques2020ecapa,ravanelli2021speechbrain,kuzmin2022magnitude}. 
We therefore propose a new model architecture combining several recent advances in deep learning together with RawNet2 to overcome this challenge.

Meanwhile, self-supervised learning has arisen as an alternative to the currently dominant supervised learning paradigm which requires human-annotated labels. 
Various frameworks in self-supervised learning train the model without ground truth labels (i.e., unlabelled), leveraging {\em pretext} tasks which can be solved by learning the inherent data properties~\cite{zhai2019s4l}. 
The prevailing methods in self-supervised learning take the form of either contrastive-~\cite{baevski2020wav2vec,chen2020simple,he2020momentum}, reconstruction-~\cite{he2021masked,bao2021beit}, or non-negative (positive pairs only)-based learning~\cite{grill2020bootstrap,caron2021emerging,baevski2022data2vec}.  

Self-supervised learning is also an actively ongoing field of research in speaker recognition~\cite{huh2020augmentation,xia2021self,mun2020unsupervised,tao2021self,sang2021self}.
However, none of these works has yet adopted an end-to-end architecture that directly operates upon raw waveform, leaving the potential open.

In this paper, we investigate various components of the raw waveform speaker recognition by exploring the aforementioned issues.
Concretely, we make the below contributions:
\vspace{-5pt}
\begin{itemize}
    \item We propose a new raw waveform speaker recognition architecture, namely RawNet3, that demonstrates EER under 1\% in the VoxCeleb1 evaluation protocol;\vspace{-5pt}
    \item We explore raw waveform speaker verification model with a self-supervised learning framework for the first time and outperform contrastive-based existing works;\vspace{-5pt}
    \item We demonstrate the effectiveness of self-supervised pre-training under semi-supervised learning scenario.
    \vspace{-5pt}
\end{itemize}

\begin{figure}[t!]
  \centering
  \includegraphics[width=0.5\linewidth]{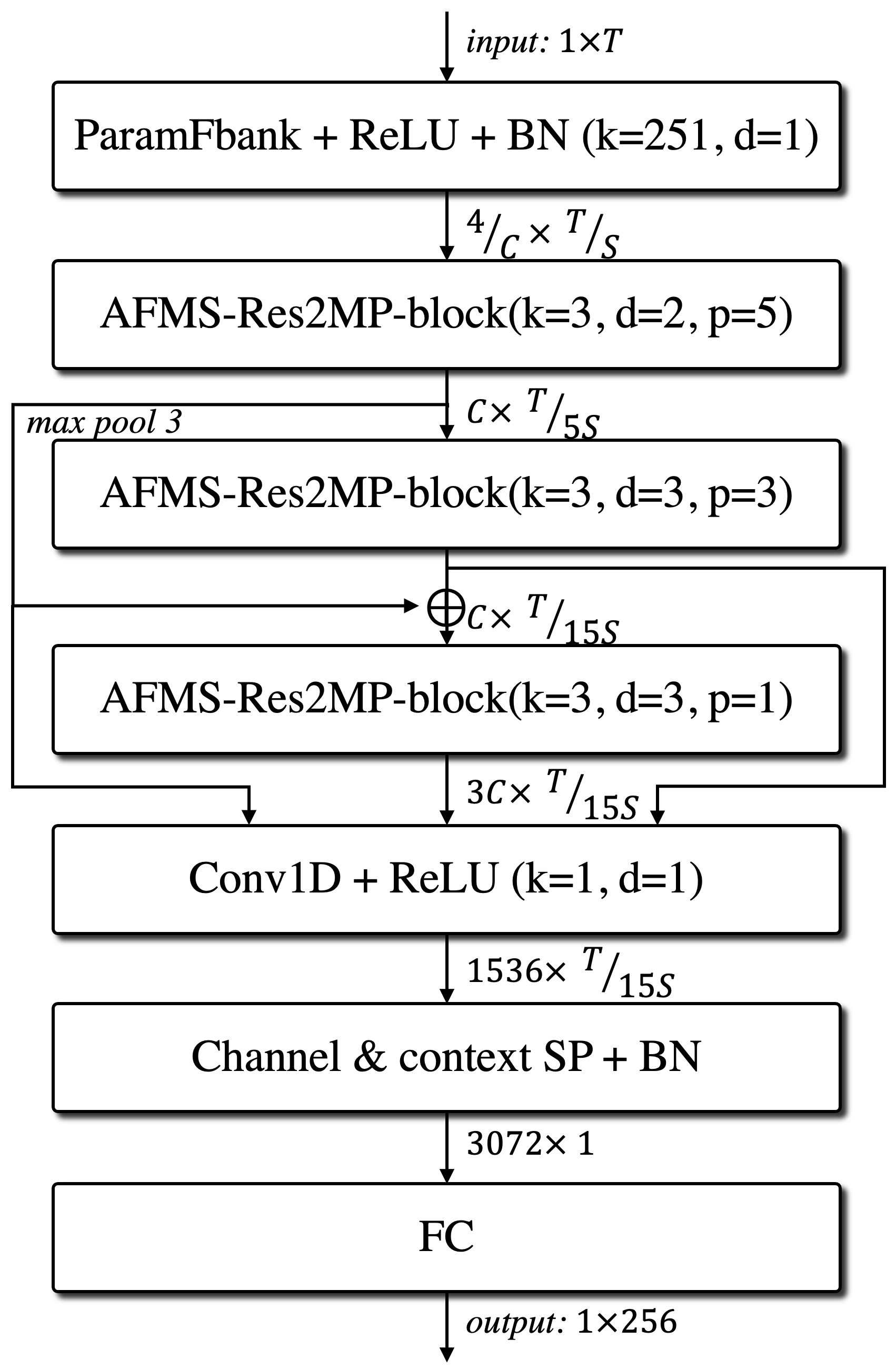}
  \caption{The RawNet3 architecture. It is in a hybrid form of the ECAPA-TDNN~\cite{desplanques2020ecapa} and the RawNet2~\cite{jung2020improved} with additional features including logarithm and normalisation. $k$, $d$, $p$, $C$, $S$, and $\oplus$ correspond to kernel length, dilation, max pooling size, number of channels, stride size of the parameterised filterbank layer, and element-wise addition.} 
  \label{fig:architecture}
  \vspace{-15pt}
\end{figure}
\begin{figure}[t!]
  \centering
  \includegraphics[width=0.5\linewidth]{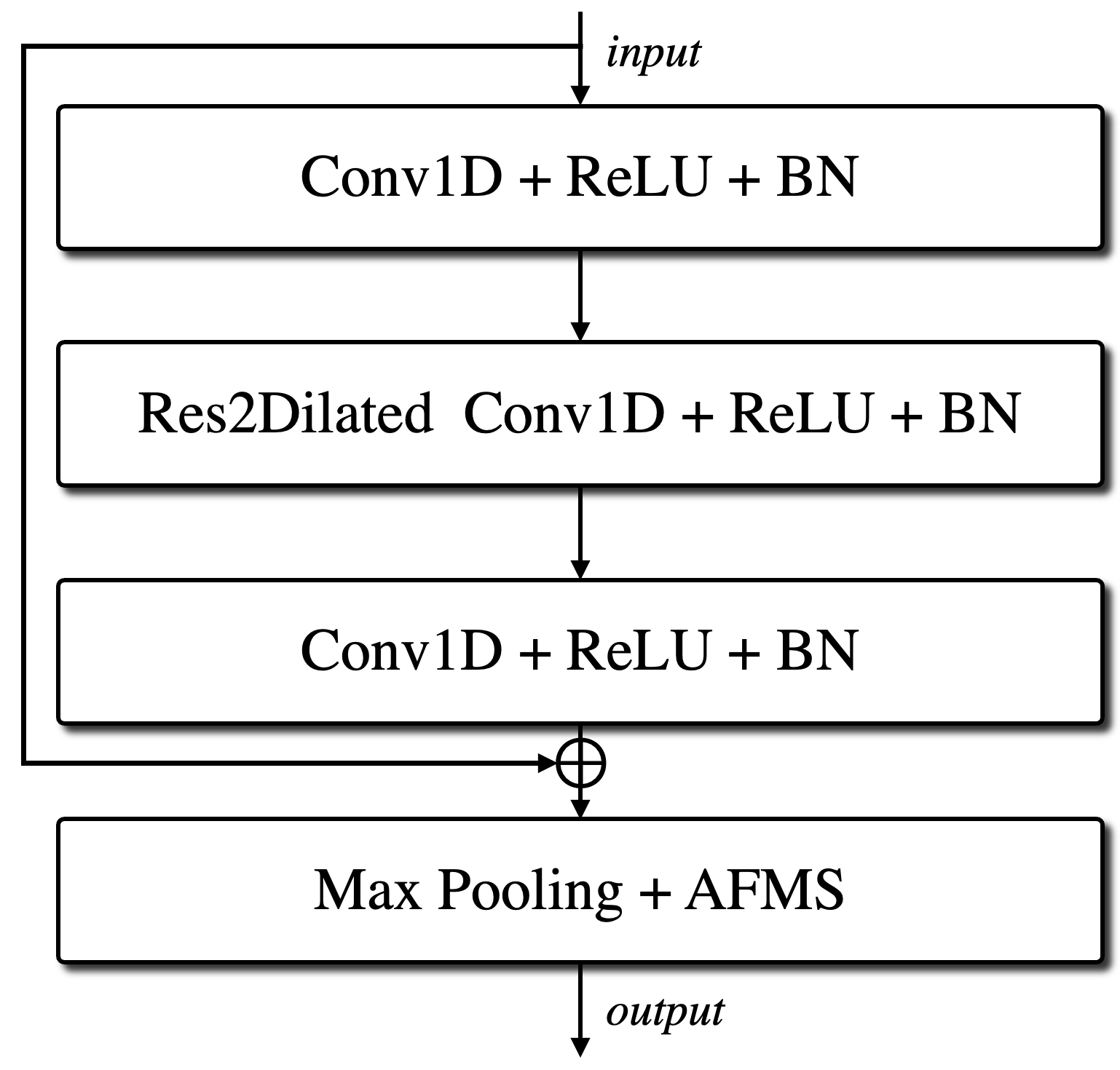}
  \caption{The AFMS-Res2MP-block of the RawNet3 architecture. AFMS refers to the extended feature map scaling module of RawNet2.}
  \vspace{-15pt}
  \label{fig:backbone}
\end{figure}

\section{Proposed architecture}
\label{sec:architecture}
Figure \ref{fig:architecture} illustrates the proposed model architecture, named {\em RawNet3} for brevity.
The architecture of RawNet3 is inspired by techniques used in
RawNet2~\cite{jung2020improved} and ECAPA-TDNN~\cite{desplanques2020ecapa}. 
We first apply pre-emphasis to the raw waveform following~\cite{jung2018complete} and feed it through an instance normalisation layer~\cite{ulyanov2016instance}.
Then the output is processed into a time-frequency representation using parameterised analytic filterbanks~\cite{pariente2020filterbank} where complex-valued filterbanks are learned. 
This layer is an extension of the sinc-convolution layer~\cite{ravanelli2018speaker}, which has been adopted in RawNet2~\cite{jung2020improved}, where real-valued parameterised filterbanks are learned.
At this step, the extent of sequence compression is configured via the stride size, where a smaller stride slows processing but produces better performance. In contrast, a bigger stride delivers faster processing but degrades performance.
The default kernel length and stride are 251 and 48, respectively, generating approximately $15$ ms window and $3$ ms shift per frame.  

Three backbone blocks with residual connections digest the parameterised analytic filterbank layer output. 
The outputs of the three blocks are concatenated identical to the ECAPA-TDNN architecture.
We also input the summation of the first and second block outputs to the third block. 
This is similar to but slightly different from the ECAPA-TDNN, which inputs all previous block outputs as the following block input.
We apply max pooling of sizes 5 and 3 to the first two backbone blocks, different to ECAPA-TDNN, following the RawNet2 architecture. 
The additional reduction of sequence length is mandatory in raw waveform speaker verification models because the sequence length is extremely longer than the handcraft feature-based model counterparts. 
The adoption of max pooling is based on our previous empirical results that it mitigates overfitting in raw waveform speaker verification models~\cite{jung2018avoiding}. 
Each backbone block, illustrated in Figure \ref{fig:backbone} and referred to as AFMS-Res2MP-block, is based on Res2Net~\cite{gao2019res2net}. It is similar to the backbone block of the ECAPA-TDNN except for two modifications: (i) AFMS from RawNet2 is applied in place of squeeze-excitation~\cite{hu2018squeeze} based on previous empirical results, and (ii) we optionally apply max pooling before applying AFMS. 

After the three backbone AFMS-Res2MP-blocks, one convolution layer with batch normalisation~\cite{ioffe2015batch} positions.
For supervised learning with classification loss (Section \ref{ssec:sl}), the classification head exists as the output layer.
For self-supervised learning with the DINO framework (Section \ref{ssec:ssl}), additional DINO head layers are added.

\vspace{-8pt}
\newpara{Comparison with RawNet2 architecture.}
Several different design choices have been adopted in RawNet3 compared to the previous RawNet2 architecture.
First, the parameterised analytic filterbank layer~\cite{pariente2020filterbank} is utilised instead of sinc-convolution layer~\cite{ravanelli2018speaker}.
Second, log and mean normalisation is applied to the analytic filterbank output.
Third, the number of backbone blocks and their connections have been adapted, following the ECAPA-TDNN alike topology. 
Last, the channel and context-dependent statistic pooling replaces a uni-directional gated recurrent unit layer~\cite{chung2014empirical}. 

\section{Adopted frameworks}
\label{sec:framework}

\subsection{Supervised learning: classification with AAM-softmax}
\label{ssec:sl}
The two most prevailing supervised learning frameworks in speaker verification are classification and metric learning. 
We adopt the classification-based training framework.
In the classification framework, a model is trained as a closed set speaker identification model with a classification head that has a dimensionality equal to the number of speakers in the train dataset (i.e., the d-vector framework~\cite{variani2014deep}).
A categorical cross-entropy objective function is adopted, which calculates the loss by comparing the classification head where softmax non-linearity is applied with the one-hot ground truth label. 
After the training is complete, the classification head is removed, and the model is used as a speaker embedding extractor.

Specifically, we adopt the AAM-softmax objective function~\cite{deng2019arcface}, also known as the ArcFace.
The AAM-softmax can enforce larger gaps between the nearest speakers.
Let $\textbf{x}_i$, $y_i$, and $\textbf{W}$ be the speaker embedding, its corresponding label, and the weight matrix of the classification head where $i$ is the index of an utterance within a mini-batch of size $N$, $0<i<N$. 
The AAM-softmax loss is formulated as:
\begin{equation}
    \mathcal{L}_A=-\frac{1}{N}\sum^{N}_{i=1}log\frac{e^{s(cos(\theta_{y_i, i})+m)}}{e^{s(cos(\theta_{y_i, i})+m)} + \sum^{N}_{j=1,j\neq y_i}e^{s \, cos(\theta_{j, i})}}
\end{equation}
where $cos(\theta_{j, i})$ is the dot product between $l2$ normalised $\textbf{x}_i$ and $\textbf{W}_j$, $s$ and $m$ are the hyper-parameters for a scale factor and a margin, respectively. 
Readers are guided to \cite{deng2019arcface} for details.

\subsection{Self-supervised learning: DINO}
\label{ssec:ssl}
The DINO framework~\cite{caron2021emerging} is one of the most competitive frameworks in the recent self-supervised learning literature. 
It compares multiple different views generated from a single utterance like the BYOL~\cite{grill2020bootstrap} framework and employs a self-distillation method similar to the data2vec~\cite{baevski2022data2vec} framework.  

DINO involves a teacher and a student network with an identical architecture but different parameters. 
Multi-crop training is utilised where a set of views ($V$), two different global ($a^g_1$ and $a^g_2$) and several local views ($a^l$), of an utterance are exploited. 
The teacher digests only global views, whereas the student digests all views.
Only the parameters of the student network are updated using the loss function. 
The parameters of the teacher network are updated as an exponential moving average of the student networks.

In the DINO framework, the outputs of both teacher and student networks are sharpened via dedicated temperatures to avoid representation collapse during training.
Centring is additionally applied to the teacher output for the same purpose, which subtracts centre from the teacher output. 
The centre is updated using an exponential moving average. 
Centring uniformly flattens the probability distribution while sharpening narrow-downs the width of the distribution.
It is argued that a good balance between centring and sharpening hyper-parameters leads to successful training without requiring architecture differentiation.
The DINO loss is then defined as:
\begin{equation}
    \mathcal{L}_D=\sum_{a\in\{a^g_1, a^g_2\}} \sum_{a'\in V,a'\neq a} H(P_t(a),P_s(a')),
\end{equation}
where $P_t$ and $P_s$ correspond to teacher and student network outputs and $H(\cdot)$ is the cross-entropy.
Readers are guided to \cite{caron2021emerging} for full details.

\begin{table}[t!]
  \caption{Results on supervised learning using the AAM-softmax~\cite{deng2019arcface} objective function. Trained on VoxCeleb1\&2 development sets. The two numbers in {\em Hz} denote frame resolutions after the first parameterised filterbank and the last max pooling layer.}
  \label{tab:SL}
  \centering
  \begin{tabularx}{\linewidth}{lYY}
    \toprule
    \textbf{Configurations} & \textbf{EER(\%)} & \textbf{minDCF}\\
    \midrule
    RawNet2~\cite{jung2020improved} & 2.48 & N/R\\
    \midrule
    RawNet3 (stride=48) & 1.05 & 0.0763\\
    \hline
    $-$ param fbank log & 1.27 & 0.0852\\
    $-$ param fbank norm & 1.22 & 0.0838\\
    $-$ param fbank log\&norm & 1.23 & 0.0927\\
    $-$ ch\&context stat pool & 1.45 & 0.0975\\
    \hline
    $\rightarrow$ stride=10, 1600Hz$\rightarrow$106Hz & \textbf{0.89} & 0.0659\\
    $\rightarrow$ stride=16, 1000Hz$\rightarrow$66Hz & 0.90 & \textbf{0.0593}\\
    $\rightarrow$ stride=24, 666Hz$\rightarrow$44Hz & 0.96 & 0.0773 \\
    $\rightarrow$ stride=64, 250Hz$\rightarrow$16Hz & 1.11 & 0.0851\\
    $\rightarrow$ stride=96, 166Hz$\rightarrow$11Hz & 1.31 &0.0937\\
    
    \bottomrule
  \end{tabularx}
  
\end{table}

\section{Experiments}
\label{sec:exp}
We present three sets of experiments using the proposed RawNet3 model under various frameworks: (i) supervised learning using ground truth label-based classification (Table~\ref{tab:SL}); (ii) self-supervised learning using the DINO framework (Table~\ref{tab:SSL}); and additionally, (iii) semi-supervised learning (Table~\ref{tab:SemiSL}). 
In the semi-supervised learning scenario, we first pre-train the model using the DINO self-supervised learning framework. 
We then fine-tune the model using supervised learning with ground truth label-based classification. 
We also provide comparison with recent models in the literature in Tables \ref{tab:literature_sl} and \ref{tab:literature_ssl}.

\subsection{Dataset}
\label{ssec:db}
We use VoxCeleb1 and 2 datasets \cite{nagrani2017voxceleb,chung2018voxceleb2} throughout this paper, which contain utterances collected in the wild, including background noises.
The two datasets respectively include utterances from $1,251$ and $6,112$ speakers and more than $340$ and $2,440$ hours of speech. 
VoxCeleb1 is divided into two subsets: the development set, which involves $1,211$ speakers and the evaluation set, which involves $40$ speakers.
VoxCeleb2 is also divided into two subsets: the development set, which involves $5,994$ speakers and the evaluation set, which involves $118$ speakers.

\subsection{Data configurations and metrics}
\label{ssec:exp_config}
For supervised learning experiments, we employ the development sets of the VoxCeleb1 and 2 datasets.
For self-supervised learning experiments, the development set of VoxCeleb2 is used.
For fine-tuning the self-supervised pre-trained model, i.e., the semi-supervised scenario, we pre-train the model using the VoxCeleb2 development set and fine-tune the model using the VoxCeleb1 development set.
The VoxCeleb1-O benchmark evaluation protocol that involves the VoxCeleb1 test set is used to measure the performance for all experiments.

The widely adopted EER (\%) is the primary metric.
We also report performances using the minimum detection cost function (minDCF) metric. 
For the minDCF, hyper-parameters of $P_{target}$=0.05 and $C_{false alarm}=C_{miss}$=1 are adopted.
Lower values depict superior performances for both metrics.

\begin{table}[t!]
  \caption{Results on self-supervised learning using the DINO~\cite{caron2021emerging} framework. Trained on VoxCeleb2 development set.}
  \label{tab:SSL}
  \centering
  \begin{tabularx}{\linewidth}{lYY}
    \toprule
    \textbf{Configurations} & \textbf{EER(\%)} & \textbf{minDCF}\\
    \midrule
    RawNet3 & \textbf{5.74} & \textbf{0.3507}\\
    \midrule
    $-$ param fbank log & 10.46 & 0.5775\\
    $-$ param fbank mean norm & 8.87 & 0.4969\\
    $-$ param fbank log\&mean norm & 9.98 & 0.5386\\
    \midrule
    $+$ DINO temp warm-up & 5.89 & 0.4004\\
    $+$ DINO last layer norm & \textbf{5.40} & \textbf{0.3396}\\
    \midrule
    $\rightarrow$ DINO T momentum 0.99 & 6.17 & 0.3987\\
    $\rightarrow$ half batch size (400$\rightarrow$200) & 6.87 & 0.4513\\
    \bottomrule
  \end{tabularx}
  
\end{table}
\begin{table}[t]
  \caption{Results on fine-tuning the pre-trained model. Trained on VoxCeleb1 development set.}
  \label{tab:SemiSL}
  \centering
  \begin{tabularx}{\linewidth}{lYY}
    \toprule
    \textbf{Configurations} & \textbf{EER(\%)} & \textbf{minDCF}\\
    \midrule
    RawNet3 (w/ pre-train) & \textbf{2.18} & \textbf{0.1519}\\
    RawNet3 (w/o pre-train) & 2.98 & 0.2268\\
    
    \bottomrule
  \end{tabularx}
  
\end{table}

\subsection{Implementation details}
\label{ssec:implementation}
\newpara{Common.} 
All experiments have been conducted using the PyTorch Python library with four Nvidia V100 GPUs.  
The model architecture, training recipe and pre-trained model weights will be freely accessible.\footnote{will be available in \url{https://github.com/Jungjee/RawNet} and \url{https://github.com/clovaai/voxceleb_trainer} prior to publication.}
Unless mentioned otherwise, the parameterised filterbank layer has a kernel length of $251$, stride size of $48$, and $256$ filters.
AFMS-Res2MP blocks have $1024$ filters, and other hyper-parameters, including kernel length, pool size, and dilation, follow that of Figure~\ref{fig:architecture}. 
The Adam optimiser~\cite{kingma2015adam} with SGDR learning rate scheduling~\cite{loshchilov2017sgdr} is adopted. 
Weight decay regularisation of $5e^{-5}$ is applied to the model.

\newpara{Supervised learning.}
The learning rate is scheduled between $1e^{-3}$ and $5e^{-6}$ and restarts every eight epochs.
The model is trained for $40$ epochs.
AAM-softmax has a margin $m$ of $0.3$ and scale $s$ of $30$.
The model is trained using randomly cropped $3$ seconds utterances. 
The size of a mini-batch is $512$. 
We apply batch augmentation, similar to that of \cite{hoffer2020augment} where augmentation methods include waveform masking, speed, noise and reverberation additions. 

\newpara{Self-supervised learning.}
The learning rate is scheduled between $1e^{-3}$ and $1e^{-5}$ and restarts every $16$ epochs. 
The model is trained for $80$ epochs.
The size of a mini-batch is $400$ unless mentioned otherwise.
We adopt two global views of $4$ seconds for the DINO framework and five local views of $2$ seconds.
Augmentation involves noise and reverberations with additional data curriculum augmentation~\cite{heo2022curriculum}. 
Temperatures for the teacher and the student RawNet3 models are $0.04$ and $0.1$, respectively.
Momentum values for the teacher model and centre update are $0.987$ and $0.9$, respectively.

\begin{table}[t]
 \caption{Comparison with recent literature of supervised speaker verification. $^{\dagger}$: calculated with $P_{target}=0.01$.}
  \centering
  \label{tab:literature_sl}
  \begin{tabularx}{\linewidth}{lYYY}
  \toprule
    & \textbf{In Feat} & \textbf{EER(\%)} & \textbf{minDCF}\\
  \toprule 
  Desplanques et al.\cite{desplanques2020ecapa} & MFCC & 0.87 & 0.1066$^\dagger$\\
  Ravanelli et al.\cite{ravanelli2021speechbrain} & Fbank & 0.69 & N/R\\
  Kuzmin et al.\cite{kuzmin2022magnitude} & Fbank & \textbf{0.66} & \textbf{0.0640}$^\dagger$\\
  \midrule
  Zhu et al.\cite{zhu2021vector}& Waveform & 2.60 & 0.2390\\
  Li et al.\cite{li2021fdn} & Waveform & 2.31 & N/R\\
  Lin et al.\cite{lin2020wav2spk} & Waveform & 1.95 & 0.2030\\
  Kim et al.\cite{kim2021rawnext} & Waveform & 1.29 & 0.1420\\
  \textit{\textbf{Ours}}$-$stride=10 & Waveform & \textbf{0.89} & 0.0659\\
  \textit{\textbf{Ours}}$-$stride=16 & Waveform & 0.90 & \textbf{0.0593}\\

  \bottomrule
  \end{tabularx}
  
\end{table}

\section{Results}
\label{sec:results}
\subsection{Supervised learning}
\label{ssec:res_sl}
Table~\ref{tab:SL} reports the result of RawNet3 architecture under supervised learning with ground truth label-based classification.
Compared to the previous RawNet2 architecture, RawNet3 demonstrates superior performance with EER reduced from $2.48\%$ to $1.05\%$, showing over $57\%$ improvement relative. 

Rows from $3$ to $6$ report ablation experiments on the RawNet3 architecture, excluding each modified component.
Exclusion of either logarithm or mean subtraction normalisation degrades the performance. 
Channel and context statistical pooling had more impact as a single component, degrading the EER to $1.45\%$ when excluded. 

The last five rows demonstrate a trade-off between computation and performance via different stride sizes in the parameterised filterbank layer. 
A smaller stride size generates a longer sequence of frames, bringing further performance improvement at the cost of more computation. 
On the other hand, a larger stride size generates a shorter sequence of frames, resulting in less computation at the cost of performance degradation.  
With a stride size of $10$, the EER is reduced to $0.89\%$.

\vspace{-5pt}
\subsection{Self-supervised learning}
\label{res:ssl}
Table~\ref{tab:SSL} delivers the result of the proposed RawNet3 architecture under the DINO self-supervised learning framework.
The result in the first row shows an EER of $5.74\%$ without utilising any ground truth label. 
Rows from $2$ to $4$ show the results of excluding logarithm, mean normalisation or both after the first parameterised filterbank layer. 
Compared to supervised learning, logarithm and mean normalisation affected the performance significantly.
We analyse that this phenomenon is related to the property of the DINO framework, which does not involve negative pairs. 
Because comparison is only made among the same utterances with different augmentations, robustness towards channel variation can be weaker, which the logarithm and mean normalisation offer.

Rows 5 and 6 show ablations by testing the officially recommended hyper-parameter changes.\footnote{\url{https://github.com/facebookresearch/dino}}
Both hyper-parameters tended to act differently than anticipated. 
Applying warm-up did not affect the performance noticeably, and normalising the last layer (no normalisation is expected to perform better) further improved the EER to 5.40\%.
The last two rows present additional ablation experiments by changing the teacher network's momentum value and halving the batch size.

\subsection{Semi-supervised learning}
\label{res_semisl}
Table~\ref{tab:SemiSL} delivers the result of RawNet3 under the semi-supervised scenario where we assume that the ground truth label only exists for a smaller VoxCeleb1 development set. 
By solely initialising the RawNet3 model with DINO pre-trained weight parameters, $25\%$ relative performance improvement was observed. 
EER of $2.18\%$ was obtained with pre-training, whereas EER was $2.98\%$ without the DINO pre-training.   

\begin{table}[t]
 \caption{Comparison with self-supervised learning models.}
  \centering
  \label{tab:literature_ssl}
  \begin{tabular}{lccc}
  \toprule
    & \textbf{Framework} & \textbf{EER(\%)} & \textbf{minDCF}\\
  \toprule 
  Huh et al.~\cite{huh2020augmentation} & AP+AAT & 8.65 & 0.4540\\
  \multirow{2}{*}{Xia et al.~\cite{xia2021self}} & MOCO+Wav- & \multirow{2}{*}{8.23} & \multirow{2}{*}{0.5900}\\
  & Aug(ProtoNCE) & & \\
  Mun et al.~\cite{mun2020unsupervised} & CEL & 8.01 & N/R\\
  Tao et al.~\cite{tao2021self} & Contrastive & 7.36 & N/R\\
  Sang et al.~\cite{sang2021self} & SSReg & 6.99 & 0.4340\\
  \textit{\textbf{Ours}} & DINO & \textbf{5.40} & \textbf{0.3396}\\
  \bottomrule
  \end{tabular}
\end{table}

\subsection{Comparison with recent literature}
\label{res_literature_compare}

\newpara{Supervised learning.}
Table~\ref{tab:literature_sl} compares the RawNet3 trained with ground truth labels with recent handcrafted feature- and raw waveform-based models\cite{desplanques2020ecapa,ravanelli2021speechbrain,kuzmin2022magnitude,zhu2021vector,li2021fdn,lin2020wav2spk,kim2021rawnext}. 
Compared to the recent state-of-the-art models, our RawNet3 shows a competitive performance of EER $0.89\%$, whereas EER of the best model~\cite{kuzmin2022magnitude} is $0.66\%$. 
In terms of the minDCF, RawNet3 has a value of $0.0593$; however, the values are incomparable due to the difference in $P_{target}$.
Our RawNet3 outperforms the previous best model among raw waveform speaker verification models by $31\%$ relative ($1.29\%$ vs $0.89\%$).

\newpara{Self-supervised learning.} 
Table~\ref{tab:literature_ssl} compares the proposed RawNet3 trained using the DINO self-supervised framework with other recent works in the literature. 
All five mentioned works~\cite{huh2020augmentation,xia2021self,mun2020unsupervised,tao2021self,sang2021self} can be seen as contrastive learning variants. 
The DINO framework, which does not adopt negative pairs, demonstrated the best performance. 

\newpara{Relation with iterative clustering.}
Iterative clustering has been proven to be effective for self-supervised speaker verification models~\cite{cai2021iterative,tao2021self}.
Our model can be viewed as the initial model of iterative clustering; thus, the model can enjoy the benefits of iterative clustering methods. 
The RawNet3 model trained using the DINO framework is hence complementary to the existing iterative clustering approaches.

\section{Conclusions}
\label{sec:conclusion}
This paper proposes RawNet3, a new raw waveform speaker verification model and evaluates it under three scenarios.
In supervised learning, the model demonstrates an EER of $0.89\%$, competitive with state-of-the-art handcrafted feature-based counterparts.
Using the DINO self-supervision framework, the model also demonstrates an EER of $5.40\%$, outperforming existing contrastive learning-based works. 
Utilising the DINO-trained model as pre-training, fine-tuning with a smaller dataset is also proven to be effective, showing $25\%$ relative improvement to the model trained with random initialisation.

\bibliographystyle{IEEEtran}
\clearpage
\bibliography{shortstrings,mybib}
\end{document}